\documentclass[twocolumn,showkeys,preprintnumbers,amsmath,amssymb,pre]{revtex4-1}

\pdfoutput=1 
\usepackage[pdftex]{graphicx}
\usepackage{dcolumn}% Align table columns on decimal point
\usepackage{bm}% bold math

\begin{document}

\title{Mechanical state, material properties and continuous description of an epithelial tissue}

\author{Isabelle Bonnet$^{1}$}
\author{Philippe Marcq$^{2}$}
\author{Floris Bosveld$^{1}$}
\author{Luc Fetler$^{2}$}
\author{Yohanns Bella\"{\i}che$^{1}$}
\email{yohanns.bellaiche@curie.fr}
\author{Fran\c{c}ois Graner$^{1}$}
\email{francois.graner@curie.fr}

\affiliation{
$^1$ Genetics and Developmental Biology, Team `Polarity, division and morphogenesis',\\
Institut Curie, UMR3215 CNRS, U934 Inserm, UPMC\\
$^2$ Physico-Chimie Curie, Institut Curie, UMR168 CNRS, UPMC \\
26 rue d'Ulm, F-75248 Paris Cedex 05 France
}

\date{May 8, 2012}

\begin{abstract}

  During development, epithelial tissues undergo extensive morphogenesis based on
  coordinated changes of cell shape and position over time.
 Continuum mechanics describes 
tissue mechanical state and shape changes 
in terms of strain and stress.
It accounts for individual cell properties using only  a few  spatially averaged material parameters.
To determine the mechanical state and parameters in  the
  \textit{Drosophila} pupa dorsal thorax epithelium,
  we sever  \textit{in vivo} the adherens junctions around a disk-shaped domain comprising typically   hundred cells. This  enables
 a direct measurement of  the strain along different orientations at once.
The amplitude and anisotropy  of the strain
 increase  during 
development.
 We also measure the stress to viscosity ratio and similarly find an increase in amplitude and anisotropy.
The relaxation time is  of order of ten seconds.
We propose a space-time, continuous model of the relaxation.
Good agreement with experimental data validates 
the description of the epithelial domain  as a continuous, linear, visco-elastic material. 
 We   discuss the relevant time and length scales.
Another material parameter, the ratio of external friction to 
internal viscosity, is estimated by fitting the initial velocity profile.
Together, our results  contribute  to quantify   forces and  displacements,  
 and their time evolution during morphogenesis.
 
\end{abstract}

\maketitle

%----------------------------------------------
\section{INTRODUCTION}
\label{itd}
%-------------------------------------------

 \begin{figure}
\includegraphics{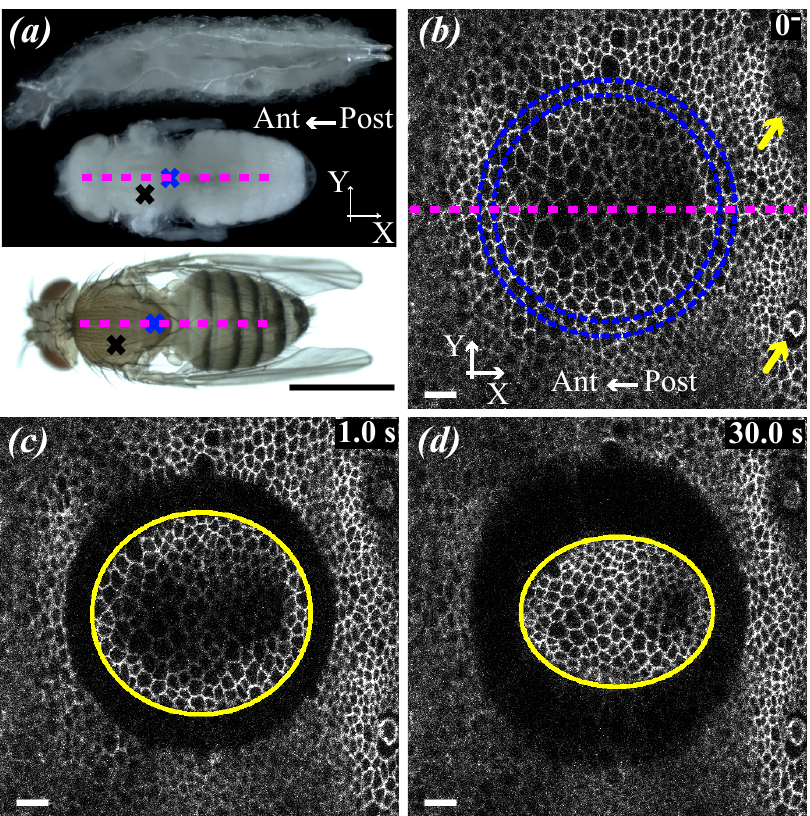}
	\caption{
 Large-size annular severing   in a fly dorsal thorax.\\
(a) Developmental stages of a fruit fly \textit{Drosophila melanogaster}.  Top: larva. Middle: pupa, with pupal case removed (ESM Figure~1) and cuticle kept intact. Bottom: adult.
Dashed lines represent the midline (symmetry axis).
The $X$  axis is antero-posterior: anterior (head) towards the left, posterior  (abdomen) towards the right. The   $Y$ axis is medio-lateral.
Crosses: approximate positions of severing, along the midline in the scutellum  (blue)  and  off-axis  in the scutum (black).\\
(b)  Epithelial cell apical junctions marked by E-Cadherin:GFP 
just before the severing, $t=0^-$, here in an old pupa 
(see text for classification).
Blue circles: two concentric circles define the annular severed region;  
the distance between circles corresponds to about  one cell size. 
Yellow arrows: macrochaete 
used as spatial references to 
position the severed region.\\
(c) First image after severing, $t=1$~s. Yellow: fitted ellipse  
\cite{thevenaz}  (see Methods). (d) Time $t=30$~s after the severing, showing a larger opening along   $Y$ than 
$X$.  Bars~$=1$ mm 
(a), $10\; \mu$m 
(b-d). 
 \label{fig:metamorphosis}
\label{fig:fit_ellipse}
}
\end{figure}

An epithelial tissue is a  
sheet of cells that acts as a barrier, separating for instance  the outside and the inside of a multicellular organism. Its biological function relies in part on the formation of a network of adherens junction belts, connected to the acto-myosin cytoskeleton, where cells adhere to each other  \cite{lecuit_lenne}, and which 
transmits mechanical information over several cell
 diameters~\cite{lecuit_lenne,Chen_2004,ma_hole_drilling}. 
A key issue is to understand and model the role of tissue mechanics 
(forces,  displacements,  and  their time evolution)
in the  coordinated changes of cell shape and position which determine morphogenetic flows at the tissue level
\cite{lecuit_lenne,lecuit_nature,mammoto2010}.  

Several models describe tissues  using   {\it continuum mechanics }   
\cite{ma_hole_drilling,fung,hufnagel,bittig_2008,blanchard,butler,ranft2010,aigouy}.
One precondition is the existence 
of a mesoscopic scale defining a domain over which averages of cell properties
are well-defined \cite{fung,text_book_meca}.
This description further relies on the assumption that the tissue mechanical state
can be quantified,
at the same mesoscopic scale,
by two variables \cite{fung}:
the stress  
characterizing 
in which directions, and to 
what extent, the domain 
is under tension or under compression; and the strain  characterizing how far the domain geometry is from that of a relaxed state.
Such description has the advantage of accounting for individual cell properties using only  a few spatially averaged parameters,  which 
 determine for instance how fast and through which succession of states the domain 
reacts to external solicitations.  

Beyond simple observation,
{\it in vivo} mechanical measurement techniques include 
elastography \cite{Ophir},
photoelasticity \cite{nienhaus}, 
magnetic micromanipulation \cite{farge}, tonometry
\cite{fleury} or nanoindentation \cite{peaucelle}. 
A large literature (for review, see {\it e.g.} \cite{rauzi_review})
has established laser ablation of individual  cell junctions 
as a tool to measure the tensions within an epithelium,
in particular during \emph{Drosophila} dorsal closure 
\cite{ma_hole_drilling,hutson2003,hutson_cavitation,hutson_modeling,kiehart, peralta07}.
This technique has allowed to measure the material relaxation time $\tau$ 
and contributed to a better characterization of morphogenetic 
processes in \textit{Drosophila} \cite{farhadifar,rauzi,landsberg,fernandez}. 

The analysis of single cell junction ablation is usually based  on 
reasonable assumptions \cite{rauzi_review,hutson2003}: that the tissue is at mechanical equilibrium just before severing;  that the ablation is effective in removing  at least part of the cell junction tension; and that during relaxation the 
velocity remains small (Reynolds number much smaller than 1).
Within these assumptions,
 the initial retraction velocity yields the value of the cell junction tension removed by the ablation,
 up to an unknown prefactor which depends on the dissipation 
 \cite{farhadifar,rauzi,landsberg,fernandez,Bosveld}.

The tensor $\sigma$ denotes the component of the stress removed by ablation. 
It arises as an average 
over several individual cell junctions within a region of the tissue.
By 
ablating straight lines in either of two perpendicular directions \cite{grill,martin} 
or by performing statistics on   single cell junction ablations in several directions  
 \cite{ma_hole_drilling}, it is possible to measure 
step by step the anisotropy of 
the ratio $\sigma/\eta$ (where $\eta$ denotes the tissue viscosity).
The stress to viscosity ratio $\sigma/\eta$ is the initial strain rate after  
the severing:  the rate of displacement that the epithelium
would spontaneously undergo if it were free.

The continuous description
applies  to other cellular materials such as foams~\cite{janiaud,Outils}.
In the present work, 
we investigate experimentally whether it also applies to epithelial tissues.
More specifically, we ask whether the epithelium strain, stress 
and material properties
can be directly measured
on a mesoscopic scale.

We address these questions in a model system, the dorsal thorax epithelium of 
\textit{Drosophila} pupa.
The  pupa is the life stage during which the larva starves, feeds upon its reserves and metamorphoses into an adult  \cite{bainbridge}.
The dorsal thorax   is a single layer of  cells
whose apical surfaces, surrounded by the adherens junction belts,  face the cuticle, which protects the pupa;  their basal surfaces face the
hemolymph, which acts as a nutrient  transporter. 
 It is composed of a large anterior region, the scutum, and a posterior tip, the scutellum. 
  The thorax, and especially the scutellum, undergoes 
extensive morphogenetic changes during the metamorphosis  
  that shapes the adult fly
 (Figure~\ref{fig:metamorphosis} (a)), making it 
a useful model to decipher the mechanisms that control tissue morphogenesis during development \cite{Bosveld}. 
It has a bilateral symmetry axis: the midline
(axis $X$, dashed line  in Figure~\ref{fig:fit_ellipse} (a-b)).
 It displays large cells, the  macrochaete 
(precursor cells of adult sensory hairs,
Figure~\ref{fig:metamorphosis} (b)), whose positions are  precisely reproducible  \cite{langevin2005}.

%----------------------------------------------
\section{ANNULAR SEVERING EXPERIMENTS}
\label{mechanical:measurements}
%-------------------------------------------

We introduce an original type of 
severing experiment. We sever by 
short pulse
laser the adherens junctions in an annular region around a $\sim$30~$\mu$m radius circular tissue domain
 (see Figure~\ref{fig:fit_ellipse} (b-d), Methods and Movies 1-3). 
The inner and outer wound margins display a comparable speed, displacement and anisotropy. Note that the same anisotropy in speed and displacement results in the outer ellipse boundary having minor and major axes orthogonal
to the minor and major axes of the inner one.
 After severing, the wound heals within tens of minutes and the pupa 
develops to adult.

Experiments are done in the scutellum 
(blue cross in Figure~\ref{fig:metamorphosis} (a)).
We classify them 
 into three groups, according to the
developmental age at which the severing is performed:
earlier than 15-16~h, approximately 18-20~h, and later than 24-26~h after pupa formation.
For brevity we call them
``young" ($N = 8$), ``middle-aged" ($N = 5$) and ``old" ($N = 10$) pupae, respectively.
Similar experiments can be performed in other locations. 
As a proof of principle, we include three
 experiments performed in the scutum 
(black cross in Figure~\ref{fig:metamorphosis} (a))
of middle-aged ($N = 2$) and old ($N = 1$) pupae.

\begin{figure}
\includegraphics{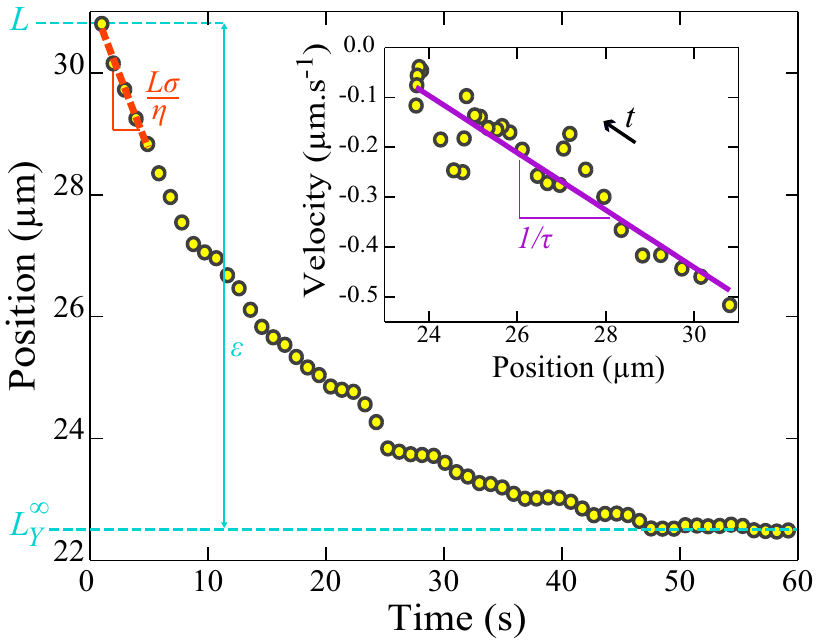}
\caption{
\label{fig:tau_x_vs_v}
Model-independent measurement of strain $\varepsilon$, stress to viscosity ratio $\sigma/\eta$, and relaxation time $\tau$.
The wound margin position
(ellipse semi-axis) is plotted  {\it versus} time after severing;
 data from Figure~\ref{fig:fit_ellipse} (b-d), 
ESM Movie 1,
along the $Y$ axis. The difference between the initial ($L$) and final ($ L_Y^\infty$) positions (blue dashed lines) 
directly yields the value of $\varepsilon = \ln  (L/L_Y^\infty)$. 
The initial velocity $- \frac{\mathrm{d}Y}{\mathrm{d}t}(t=0) $
(slope of the orange dashed line) yields an estimate of $ L \sigma/\eta$.
Inset: Velocity, estimated by finite differences of successive positions, 
{\it versus} the position during the first $30$ s. 
An arrow indicates the direction of increasing time $t$.  
The slope of a linear fit (purple line) yields the inverse of 
the relaxation time, $1/\tau$ (see Equation~\ref{eq:measure_tau}).
  }
\end{figure}

We analyse the retraction of the inner tissue domain and its margin,
within the set of assumptions relevant for the retraction of a single 
severed cell-junction.
The boundary of the retracting domain 
is fitted at each time point by an ellipse  \cite{thevenaz}  (see Methods
and Figure~\ref{fig:fit_ellipse} (c-d)),
yielding values of the $X$ and $Y$ axes as a function of time (Figure~\ref{fig:tau_x_vs_v}).

 Our protocol offers several advantages.
(i) The experiment directly yields   measurements averaged over
the severed tissue domain, comprising typically a hundred of cells.
As discussed below, the
 domain size is thus an adequate mesoscopic scale,
crossing-over from
 detailed cell-level to tissue-level continuous descriptions \cite{fung,text_book_meca}. 
(ii)  The comparison between the initial and final states yields a direct measurement of the strain which existed before severing. 
(iii) Several cell-cell junctions in various directions are severed at once (rather than step by step) and in the same pupa. Hence
 one experiment yields a  direct measurement of all  
 stress to viscosity tensor components. 
 It takes only a few tens of seconds,
the typical value of the relaxation time.
(iv) Last but not least, by isolating in time and space  an epithelium
 domain from its  neighboring cells, we obtain a complete set of 
well-defined  initial, boundary and final conditions for a spatio-temporal model  of the retracting domain. 
We thus introduce a continuous model, in the spirit
of Mayer {\it et al.}~\cite{grill} and taking into account a coupling between space and time dependences.
As explained below,  fitting such a    model   to the displacements in the domain  bulk
tests the continuous description, suggests an interpretation of the relaxation time, and
probes the  competition between internal viscosity and external friction. 

%----------------------------------------------
\section{RESULTS}
\label{results:relaxation}
%-------------------------------------------

\subsection{Model-independent measurement
of strain, stress to viscosity ratio and relaxation time}

Severing the epithelium reveals the displacement  that the epithelium
would spontaneously undergo if it were free. 
We measure the strain using a logarithmic definition 
\cite{janiaud,hoger,farahani},  Hencky's ``true strain" \cite{tanner2003}:
\begin{equation}
  \label{eq:def:hencky}
 \varepsilon_{XX} = \int_{L_X^\infty}^L \frac{\mathrm{d}X}{X} 
= \ln  \left( \frac{L}{L_X^\infty} \right), 
\end{equation}
where $L$ is the initial radius, and
$L_X^\infty$ is the ellipse semi-axis  at the end of the relaxation. 
Similarly $\varepsilon_{YY} = \ln  (L/L_Y^\infty)$. 
The advantages of this definition are that \cite{janiaud,Outils}: it is valid at all amplitudes;
it is adapted for tensorial measurements; 
and finally it increases the range of validity of the linear elasticity
approximation which applies here even to the highest strain value we measure, $0.45$.

\begin{figure}
\includegraphics{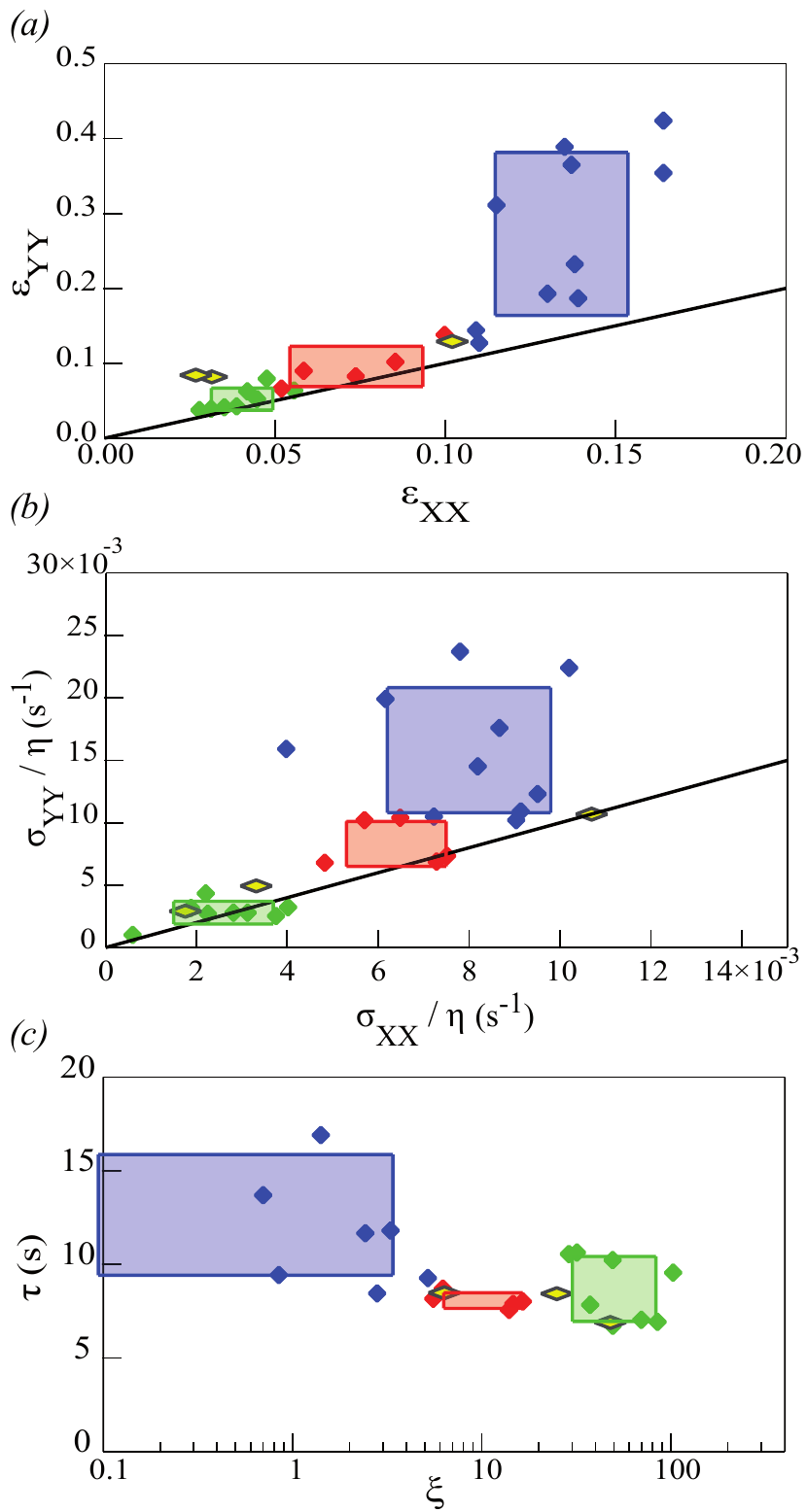}
\caption{
\label{fig:strain_anisotropy}
\label{fig:sigma_anisotropy}
\label{tau_vs_logxi}
Mechanical state and material properties.  
Color code according to pupa development ages: 
green, young; red,  middle-aged; blue, old. 
 The edges of the rectangular regions represent the mean values 
$\pm$ standard deviations for each  group.
 The experiments in the scutum are in yellow. 
(a) 
Strain anisotropy: $\varepsilon_{YY}$ {\it versus} $\varepsilon_{XX}$.
Note the difference in horizontal and vertical scales; the solid line is the  first bisectrix $Y=X$, indicating the reference for isotropy. 
(b) 
 Same for the severed stress to viscosity ratio 
$\sigma / \eta$.
(c) Relaxation time $\tau$  and  dimensionless friction to viscosity ratio $ \xi$; values are the averages of the measurements along the $X$ and $Y$ axes (ESM Figure~3).
Note the semi-log scale.
The blue rectangle takes into account two very small values of $\xi$, of order  $10^{-3}$ and $10^{-4}$ (below the plotted range).
}
\end{figure}

 Each severing is followed by a relaxation to a final  domain strictly smaller than the initial disk 
 (Figure~\ref{fig:fit_ellipse} (d)),
indicating that  before the severing the tissue had a  positive strain 
in all directions. 
Since the midline is a symmetry axis, we expect that the strain axes are parallel and perpendicular  to it.
We check that this is the case for the fitted ellipse axes, and that accordingly the shear strain $\varepsilon_{XY}$  is  indistinguishable from 0. 
 We thus plot 
 $\varepsilon_{XX}$ and $\varepsilon_{YY}$ (Figure~\ref{fig:strain_anisotropy} (a)), 
obtained with absolute precision better than $10^{-2}$.
The values are clustered according to the three pupa age groups:
 $\varepsilon$ is low and isotropic at young age (green cluster), moderate and isotropic at middle age (red), 
high
and anisotropic 
($\varepsilon_{YY}  >  \varepsilon_{XX} $) 
at old age (blue). 

Similarly, in each experiment, the initial retraction of the wound margins
 (Figure~\ref{fig:fit_ellipse} (c)) indicates the sign of the stress:  before the severing the tissue was under tensile stress in all directions. 
 This is reminiscent of positive cell junction tensions observed 
in the wing \cite{farhadifar} and in the notum \cite{Bosveld}. 
The initial retraction velocity divided by $L$ yields 
the stress to viscosity ratio $\sigma/\eta$ (Figure~\ref{fig:tau_x_vs_v}), in a way which is likely 
to be
independent of any  tissue rheological model,
as suggested by the following argument.
 When ablating a single cell-cell junction, force balance shows that the force $F$ exerted by the ablated junction on the neighbouring vertex is proportional to the 
initial
recoil velocity $v$: $F=-\gamma v$, where $\gamma$ is a friction coefficient \cite{rauzi_review}. Coarse-graining this relation, and  assuming that the main source of dissipation is identical in our  experiments which sever many junctions at once, yields $\sigma \sim \eta v/L$, where the factor $L$ is introduced by dimensional analysis. 
 Results for 
the stress to viscosity ratio
$\sigma/\eta$ (Figure~\ref{fig:sigma_anisotropy} (b)) are similar to those for $\varepsilon$:
 its amplitude and anisotropy   are initially small and increase with age;
 the values are clustered according to the three age groups;
no shear stress is detected.
Values of $\sigma/\eta$ span  almost two decades, 
reflecting the sensitivity and precision of the method. 

The relaxation time, $\tau$,
is obtained from the time evolution of the ellipse axes.
Instead of fitting  an exponential to the data, 
we plot the velocity {\it versus}  position 
(Inset of Figure~\ref{fig:tau_x_vs_v}).
This method is robust and  independent of any assumption.
A linear regression of the velocity as a function of position
yields a slope of $-1/\tau$, as seen, \emph{e.g.}, in the equation:
\begin{equation}
 \label{eq:measure_tau}
\frac{\mathrm{d}Y}{\mathrm{d}t}(t)  = - \frac{Y(t) - L_Y^{\infty}}{\tau_Y}.
\end{equation}
We find that  the relaxation time is approximately isotropic
    (ESM Figure~3 (A)) and use $ (\tau_X +\tau_Y)/2$ as an estimate of $\tau$.
    It is close to 10 s and only slightly varies with the pupa age group 
(Figure~\ref{tau_vs_logxi} (c), vertical axis).

\subsection{Space-time model and measurement of the friction to viscosity ratio}

\begin{figure*}
\includegraphics{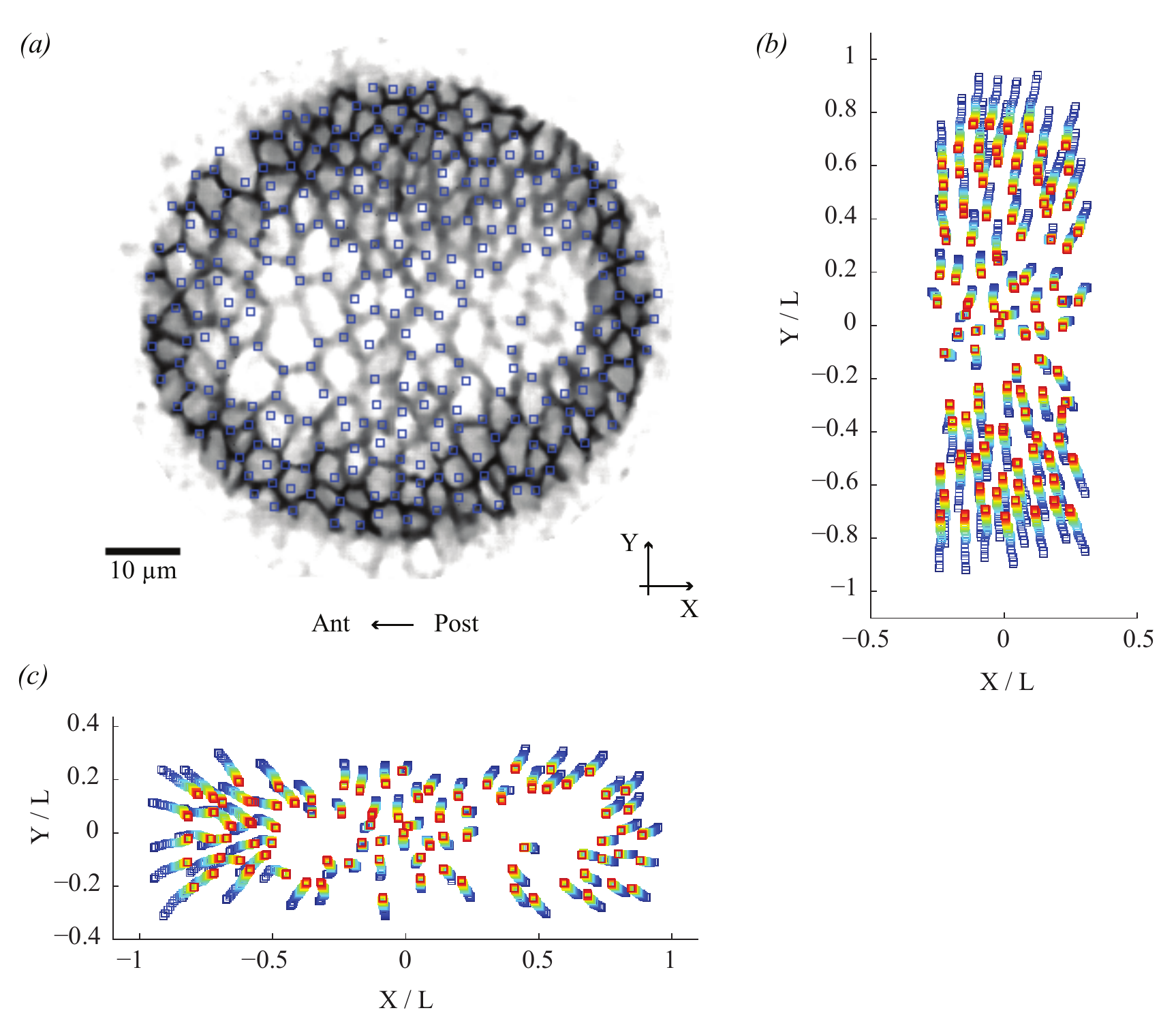}
\caption{\label{fig:klt}
 Tracking of  feature positions.
(a) Positions of features (blue squares), old pupa, data of 
Figure~\ref{fig:fit_ellipse} (c) after image denoising. 
The logarithms of intensity levels are represented
in grey scale, and contrast is inverted for clarity. 
(b) Tracking   for the first 30 s (time color-coded from blue to red), 
inside a rectangle along the $Y$ axis. 
(c) Same for the $X$ axis.}
\end{figure*}

\begin{figure}
\includegraphics{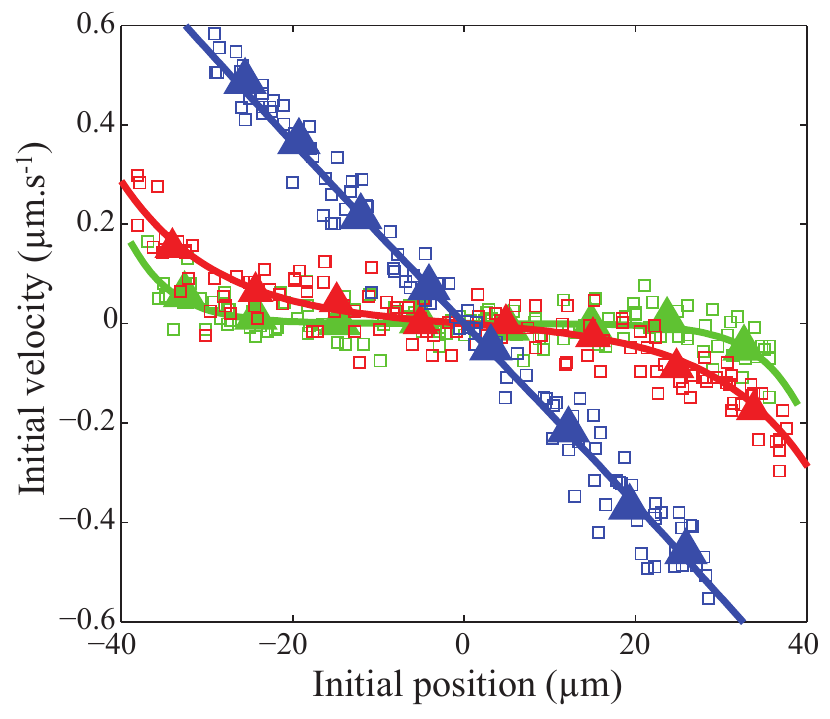}
\caption{
\label{fig:fit:xi}
Model-dependent measurement of the friction to viscosity ratio $\xi$.
Initial velocity profiles 
  immediately after severing are plotted
 {\it versus} initial position prior to severing, for three typical pupae: young (green), middle-aged (red) and old (blue). Open squares: features, from 
Figure~\ref{fig:klt} (b). Closed triangles: same, spatially averaged in 8 bins.
Lines: fit by a sinh  function (Equation~\ref{eq:vinit}), yielding  for    $\xi$ a value either  above, in, or below the measurable range:
$\xi>60$ (green),  $\xi \sim 14 \pm$5 (red), $\xi <0.5$ (blue), respectively.
}
\end{figure}

To describe the bulk of the retracting domain, we track the displacements of around a hundred features 
in a band of tissue
(Figure~\ref{fig:klt} and Methods).
Spatially averaging their positions by binning them into 8 equal-size 
bins 
improves the signal to noise ratio.
We obtain the initial velocity profile {\it versus} position (Figure~\ref{fig:fit:xi}).
For old pupae, it is spatially linear; for   middle-aged pupae it is spatially nonlinear;
in the young pupae, immediately after severing only the boundaries move significantly.

To account for these observations, we formulate a 
spatio-temporal, viscoelastic, Kelvin-Voigt model 
in one dimension of space, where the coordinate $z$ denotes either $X$ or $Y$,
and $t$ denotes time
after severing (see ESM).
Thanks to the definition (\ref{eq:def:hencky}), the strain 
$\varepsilon(z,t)$ is related without approximation to the velocity $v(z,t)$ 
through:
\begin{equation}
  \label{eq:velocity_strain}
  \frac{\partial v}{\partial z} = \frac{\partial \varepsilon}{\partial t},
\end{equation}
We model  the effect of friction of the epithelium against the
hemolymph and the cuticle as an external fluid friction 
with
coefficient $\zeta$.
We find that when internal viscosity dominates, the strain remains uniform and all parts of the tissue relax exponentially with the same viscoelastic relaxation time 
$\tau = \eta/E$
where  $E$ is the   Young modulus
(ESM Figure~4 (A)). 
 When external friction dominates, strain diffuses from the boundaries 
 with a diffusion coefficient  
 $D = E / \zeta$
 and  remains thus inhomogeneous at times $t$ 
smaller than  $\tau_D= L^2/D = \zeta L^2/E$
 (ESM Figure~4 (B)).
In the general case,  the dynamical equation for the strain field $ \varepsilon(z,t)$ reads: 
\begin{equation}
  \label{eq:general:D}
 \frac{\partial \varepsilon}{\partial t} = D \;
\frac{\partial^2 }{\partial z^2}  
\left(\varepsilon + \tau \; \frac{\partial \varepsilon }{\partial t}  
\right),
\end{equation}
with 
the 
conditions $\varepsilon(z,t=0)=\varepsilon_0$ initially,
$\varepsilon(z=\pm L,t)=\varepsilon_0\exp(-t/\tau)$ at boundaries, and $\varepsilon(z,t\to \infty)=0$ at the end.
The relative importance of external friction {\it versus}
internal viscosity is quantified by   the dimensionless parameter:
\begin{equation}
  \label{eq:def:xibar}
\xi = \frac{\zeta L^2}{\eta} = \frac{\tau_D}{\tau}.
\end{equation}
Solving  Equation~\ref{eq:general:D}
at short time
and integrating over space 
 Equation~\ref{eq:velocity_strain}
at $t=0$
yields 
 the initial velocity profile:
\begin{equation}
  \label{eq:vinit}
  v(z, t=0) = - \varepsilon_0 \; \frac{L}{\tau} \;
\frac{1}{\xi^{1/2} \cosh \xi^{1/2}} \;
\sinh \left( \xi^{1/2} \frac{z}{L} \right),
\end{equation}
whose
curvature is controlled by  $\xi$.
We thus  estimate $\xi$ by  fitting  a  hyperbolic sine to experimental initial velocity profiles  
  (Figure~\ref{fig:fit:xi}).
 As for the relaxation time, we check that the
values $\xi_X$ and $\xi_Y$ determined  in the directions $X$ and $Y$
  are roughly consistent  (ESM Figure~3 (B)):
  we thus use $ (\xi_X +\xi_Y)/2$ as an estimate of $\xi$.

We find that $\xi$ varies more than $\tau$ (Figure~\ref{tau_vs_logxi} (c), 
horizontal axis). 
The middle aged group is characterized by intermediate values of 
$\xi$  (Figure~\ref{fig:fit:xi}, red), of order of 10.
According to  Equation~\ref{eq:def:xibar}, this indicates a possible effect of the external friction 
on the relaxation for scales larger than 
$L/\sqrt{\xi} \sim 10 \;\mu$m. 
For both other age groups, 
the values of $\xi$ 
should be considered 
as bounds on the order of magnitude. 
For old pupae we find  $\xi $  smaller than a few times unity (Figure~\ref{fig:fit:xi}, blue). For young
pupae, we find  $\xi $ larger than a few tens or hundred (Figure~\ref{fig:fit:xi}, green).

\begin{figure}[t]
\includegraphics{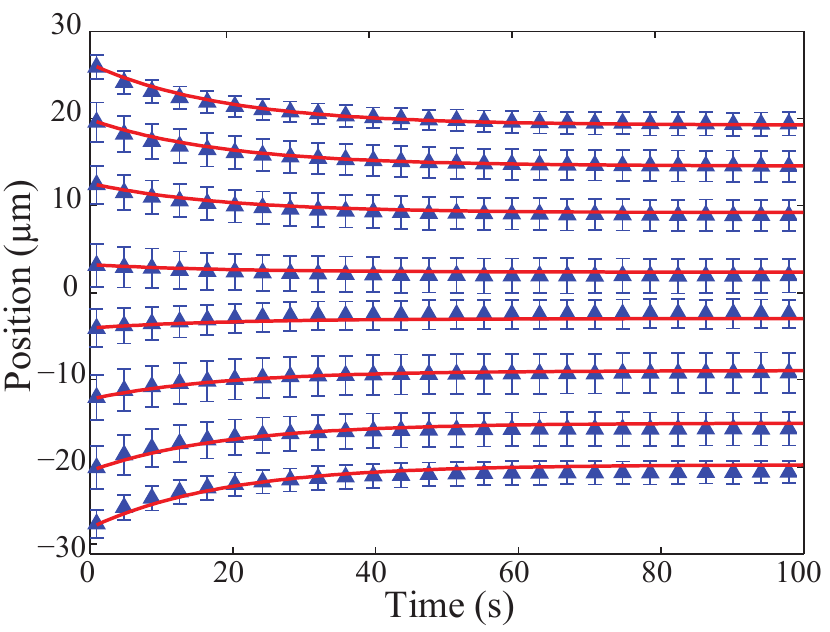}
\caption{\label{boxtraj}
Validation of the model. Blue triangles: positions of features from Figure~\ref{fig:klt} (b), spatially averaged in 
bins
{\it versus} time after severing;
bars:  standard deviation.
Red lines: numerical resolution of Equation~3, using the measured values ${\xi} = 0.08$,  $\tau = 15.8 \;s$.}
\end{figure}

This model depends on three parameters 
measured independently of each other:
the initial strain $\varepsilon_0$,
the viscoelastic time 
$\tau$, and the friction to viscosity ratio
$\xi$. 
To validate the model, we use it to numerically compute the whole space-time map of strains and displacements. We simulate each experiment using the values of $\tau$ and $\xi$  measured as described above. 
Figure~\ref{boxtraj}  
 shows that we find a good agreement with
experimental data.
Moreover, 
we plot the square deviation between
experimental   and numerical data
as a function of the parameter values $\left(\tau, \xi \right)$ 
(see Methods). 
We check that the error landscapes we obtain 
are   broadly consistent with our experimental estimates of
$\xi$ and $\tau$ (ESM Figure~5).

%------------------------------------------------------------------------------------
\section{DISCUSSION}
\label{disc_conc}
%------------------------------------------------------------------------------------

\subsection{Time scale}

We report relaxation times $\tau$  in the range 5 to 20 s, and initial 
velocities in the range 0.1 to 0.6 $\mu$m.s$^{-1}$. This agrees with values 
provided by single cell junction ablation: 10-100 s for the relaxation times, 
0.1 to 1  $\mu$m.s$^{-1}$ for the initial velocities in other 
{\it Drosophila} tissues
\cite{ma_hole_drilling,farhadifar,rauzi,landsberg,fernandez}.
The model allows to interpret $\tau$ as a viscoelastic time. 
Its  order of magnitude  is typical of the viscoelastic times 
corresponding to cell {\it internal} degrees of freedom \cite{Wottawah2005},
also measured in cell aggregates \cite{Marmottant2009,guevorkian2011}.

Conversely {\it external} degrees of 
freedom (such as cell-cell rearrangements or cell divisions) would have much larger
viscoelastic times:
in cell aggregates they are typically of order of several hours 
\cite{ranft2010,Marmottant2009,Guevorkian2010}.
The time scales of morphogenesis in this tissue are  also of order of 
hours \cite{Bosveld}, as observed in other tissues as well, such as
the  {\it Drosophila} pupal wing \cite{aigouy}.
Indeed continuous models of morphogenetic processes presuppose the existence
of such a separation of time scales.  

The present experimental data are well described assuming that a single viscoelastic time is relevant. 
Other experiments display a wide  distribution of  viscoelastic times   \cite{ma_hole_drilling,fernandez}:
the model can easily be modified   so as to take this feature into account.
Note that in the presence of friction the model  already predicts that different parts of 
the severed domain relax at different rates (ESM Figure~4 (B)). 

\subsection{Length scale}

Measurements of  $\varepsilon$, $\sigma/\eta$, $\tau$ and $\xi$ are performed as averages over the severed disk.
The requirements  on the averaging scale are that it should be (i) large enough to obtain a sufficient signal to noise ratio yielding significant results, and (ii) small enough to be able to focus simultaneously on the whole severed ring despite the epithelium surface curvature which is observed in old pupae. 
The $Y \to -Y$ symmetry  with respect to the midline observed in 
Figures~\ref{fig:fit_ellipse} (b-d), \ref{fig:fit:xi},
and \ref{boxtraj} acts as a control of the signal to noise ratio. 
The experimentally available range of scales  corresponds to 60-100 cell areas,
in other words the semi-axis length $L$ corresponds to 4-5 cell diameters. 
For old pupae, scales smaller than $L$ are also probed,
corresponding to the scale of bins (at least $L/2$ or  $L/4$,
see Figs. \ref{fig:fit:xi} and \ref{boxtraj}).

\subsection{Friction}

We experimentally determine 
the
order of magnitude of the   
dimensionless friction to viscosity ratio,  $\xi$ (Figure~\ref{tau_vs_logxi} (c), horizontal axis).
Obtaining more precise values of $\xi$ would depend on model ingredients,  such as 
depth of ablation along the apico-basal axis, detailed cell geometry, tissue
compressibility, or type of friction.

For simplicity we have modeled by a fluid friction \cite{grill}
the dissipation on either surfaces of the epithelium, namely the  hemolymph
at basal side and the cuticle at apical side.
We do not observe any threshold effect due to solid friction, 
and the agreement between model and data justifies {\it a posteriori} 
our assumption.

For young pupae, the initial velocity field is not significantly higher 
than the noise level  (Figure~\ref{fig:fit:xi}, green). In this age group, it
cannot be used to validate the  model, and the
estimate of $\xi$ strongly depends on the motion of domain parts 
located close to the boundary  (Equation~\ref{eq:vinit}).

\subsection{Continuous description}

In the dorsal thorax, the severed domain 
fulfills the theoretical requirements for  a continuous description: 
it contains a number of cells much larger than one
(large enough to allow for averaging),
while being much smaller   than the total
number of cells in the whole tissue. 

One could ask whether averages at scale $L$ are meaningful, due to variation of cell properties from cell to cell.
We speculate that if heterogeneities within the domain dominated the average,  
 the domain boundary could in principle 
  adopt any arbitrary shape compatible with the $Y \to -Y$ symmetry. 
  However, in practice, we observe that the domain boundary at the beginning,
as well as at the end, of the relaxation also displays a $X \to -X$ symmetry 
(Figure~\ref{fig:fit_ellipse} (b-d) and Movies 1-3) 
and can be fitted by an ellipse:
 this is consistent with the measurements of  $\sigma/\eta$ 
and $\varepsilon$ as tensors averaged over scale $L$.

\subsection{Model}

Limits of our model lie in its simplifications:  one dimensional
treatment, linear elasticity and viscosity, isotropy and homogeneity 
of material parameters including friction,
 identical properties of all cells contained in the severed region. 
However, the agreement between model and experiment is already 
good, and validates these approximations \emph{a posteriori}.
We emphasize that one major goal of this model
is to provide an order of magnitude of 
the friction to viscosity ratio $\xi$. 
We do not claim that this model is unique, nor that it is general.
In the future we hope to further test the model ingredients and relevance on this epithelium and other quasi-planar ones.

Since its ingredients are general, we expect our model to be of broad relevance
when studying the relaxation of quasi-planar epithelia over similar
time and length scales.
However, the simplifications involved deserve further comments. 
In fact, completely different hypotheses could also be invoked to explain 
the spatial non-linearity in the velocity profile observed for young and 
middle-aged pupae. For instance, the viscous component of biological 
tissues could be a non-linear function of time, strain, and strain rate  
\cite{fung}. A strain-dependent viscosity coefficient may lead to a 
non-linear velocity profile without the need to invoke any external friction. 
Alternatively, a spatially heterogeneous friction might also explain 
the observed velocity profiles. 
Since a linear velocity profile is observed in older pupae, 
the amplitude of these additional, more complex ingredients 
would have to decrease strongly during development.

We believe that our model has the merit of simplicity. It successfully 
accounts for the description of the velocity profiles at any time. 
Each of its hypotheses could be tested experimentally and, 
if experimental results require it, hypotheses could be relaxed to 
lead to more detailed explorations. We expect that additional ingredients 
could strongly affect the value of $\xi$, but that 
other measurements presented here 
(namely $\varepsilon$, $\sigma/\eta$ and $\tau$)
should be robust. The model thus provides a 
flexible and robust route to a mechanical description of quasi-planar epithelia.

\subsection{Perspectives}

Building upon classical single-junction laser ablations, our 
experiments enable to measure  the time evolution of 
the mechanical state (strain, and stress to viscosity ratio) and material properties (relaxation time, and friction to viscosity ratio) during pupa metamorphosis, as well as 
 their anisotropy.
This provides relevant ingredients for 
the modeling   of a tissue as a continuous,  linear,  
viscoelastic material. 
 
Moreover, our approach provides a method to explore the biological causality of the observed properties, by analysing mutant phenotypes or by micro-injecting drugs. We conjecture that  $\tau$ would be affected by low doses of drugs known to modify cytoskeletal rheology such as Cytochalasin D \cite{Flanagan}, CK-666 \cite{Nolen},Y-27632 \cite{Uehata} which respectively inhibit actin polymerization, Arp2/3 complex activity or Rho-Kinase activity. We anticipate that modulating cell-cell adhesion by knock-down of E-Cadherin or 
Catenins
 \cite{Papusheva} could affect the viscosity. Although we cannot modulate the cell-cuticle adhesion, we could  tune cell-matrix adhesion by knock-down of integrin or its linker to the cytoskeleton \cite{Papusheva} to affect the friction and thereby modify $\xi$.

In principle, by establishing a complete map of $\sigma$ at different positions, one could determine the length scale at which $\sigma$ varies within the tissue, and check {\it a posteriori} whether it is larger than the scale $L$ used to measure $\sigma$. 
However, the present experiments measure the stress up to a dissipative prefactor, 
the tissue viscosity.
Similarly, single cell-junction ablation experiments measure force up to a 
prefactor (a friction coefficient), rarely discussed or measured 
in the literature.
To separate the stress and viscosity variations, alternative methods to directly measure forces and viscosities in live epithelia are thus called for.
These include the measurement of cell junction tensions from movies observations
\cite{Brodland2010}, where the  unknown prefactor, namely the average tension over the whole image, is by definition the same for all junctions. External mechanical manipulation \cite{farge,fleury,peaucelle} yields direct determination of out-of-plane elastic and/or viscous moduli, but the in-plane moduli are only indirectly determined.  
Another possible strategy could rely on  microrheology  
\cite{MacKintosh,Panorchan,Crocker2007}: 
either passive, by tracking the brownian diffusion of beads within the cells; or active, by moving  (magnetically or optically) a bead back and forth within a cell at a specified frequency.
Such measurements could yield a direct access to the value of  $\eta$, but at the intracellular level. 
Measuring $\eta$ at the scale of several cells either directly, or indirectly by measuring $E$ and $\tau$ and using the relation $\eta=E\tau$, could probably be possible, but we are not aware yet of a published method. 

Since $\varepsilon$ is   a dimensionless geometrical quantity, 
its value is insensitive to cell size:
the observed $\sim$7-fold increase in $\varepsilon$ cannot be related to
 changes of cell size over time. It should involve other causes, 
that remain to be determined.
Part of the measured increase in  $\sigma/\eta$ values may 
be due to cell divisions which decrease the average cell area  by a factor of  
$\sim 2$ between young and old pupae.
The observed increase is a factor of $\sim 25$, and
 thus should involve other causes as well.
One candidate is a decrease of the tissue viscosity $\eta$.
However the near constancy of the viscoelastic time would then imply
a corresponding increase in tissue elasticity, a rather unlikely
coincidence. We therefore expect that stronger
tensile forces are at work in the tissue in older pupae.
From 18 hours APF onwards,
the lateral part of the scutellum undergoes apical cell 
contractions that shape the lateral domain of the tissue \cite{Bosveld}. 
This observation led us to divide the ablation experiments into the three 
age groups. It may also explain the changes in mechanical 
properties measured in the tissue for middle-aged and old pupae. 
To precisely analyse the causes of the increases in  $\varepsilon$  
and $\sigma/\eta$ would therefore require to identify genes that 
specifically affect the lateral contraction of the tissue.

Our experimental method should be applicable to other tissues.
Performing experiments in an epithelium which is flat enough to 
significantly increase $L$, and with a sufficient signal to noise ratio 
on the estimate of $\xi$, would enable to  check 
the scaling of $\xi$ {\it versus} $L$ given by Equation~\ref{eq:def:xibar}.

%-----------------------------------
\section{METHODS}
\label{methods}
%-----------------------------------
\subsection{Experiments}
{\it Drosophila melanogaster} larvae were collected at the beginning of metamorphosis (pupa formation)  \cite{droso}.
 The pupae were kept   at 25$^{\circ}$C, then dissected and mounted as described in \cite{segalen2010} (ESM Figure~1).  
 Using the adherens junction protein E-Cadherin fused to Green Fluorescent Protein    \cite{Oda2001} we 
imaged
the apical cell junctions by fluorescent microscopy
   \cite{segalen2010}.
   
The time-lapse laser-scanning microscope LSM710 NLO (Carl Zeiss MicroImaging) 
collected   512$\times$512 pixel images in  mono-photon mode at 488 nm 
excitation (pixel size in the range 0.24 to 0.29 $\mu$m) through a  
Plan-Apochromat 63$\times$1.40 oil  objective. 
Before and after severing, the time interval between two consecutive frames was either 393 ms or 970 ms
according to the scanning mode
(bi- or mono-directional), without any apparent effect on the results presented here.
We focused on the adherens junctions to be severed; due to the epithelial surface curvature especially in old pupae, adherens junctions of innermost and outermost regions were slightly out of focus (Figure~\ref{fig:fit_ellipse} and Movies 1-3).

We  defined a region of interest  as an annular region  between two concentric circles.
We severed it at the 10$^{\rm th}$ frame
(Figure~\ref{fig:fit_ellipse}) and defined $L$ as the radius of its inner circle.
Laser severing was performed 
using Ti:Sapphire laser  (Mai Tai DeepSee, Spectra Physics)   in  two-photon mode at 
890~nm,  $<$~100~fs pulses, 80~MHz repetition rate, 
$\sim$ 0.2~W at the back focal plane,
used at slightly less than full power to avoid cavitation (ESM Figure~2) \cite{hutson_cavitation}.
Severing itself had a duration ranging from 217 to 1300 ms (according to the  
size  of the region of interest  
and the scanning mode)
during which no image was acquired.
We could check that the severing had been effective to remove their adherens junctions, rather than simply bleach them
(ESM Figure~2):
 after retraction the cells at the wound margin moved apart more than a cell diameter; moreover, 
 there was no fluorescence recovery (Figure~\ref{fig:fit_ellipse} (d)).

\subsection{Analysis}
At the tissue scale,  for each frame after the severing the inner severed tissue boundary was fitted with an ellipse (Figure~\ref{fig:fit_ellipse} (c-d))  using  the
ovuscule  \cite{thevenaz} ImageJ plugin  \cite{url_ovuscule}. We have  modified it  so that each fitting procedure started from the ellipse fitted on the preceding image: such tracking improved the speed and robustness. 
Data of Figure~\ref{fig:tau_x_vs_v} are linearly fitted over a 5 frames sliding window in order to improve the signal to noise of position and velocity  determinations. 

Independently, at the cell scale, images were denoised with Safir software 
\cite{kervrann,boulanger}.
We took the logarithm of the grey levels, to obtain comparable intensity gradients in differently contrasted parts of the image (Figure~\ref{fig:klt} (a)). This allowed us to use Kanade-Lucas-Tomasi (KLT)  
\cite{klt1,klt2} tracking algorithm \cite{clemson} by selecting $\sim 10^2$ ``features" of interest,
 corresponding to most cell vertices   (Figure~\ref{fig:klt} (a)). 
Features were tracked from frame to frame: they  followed the cell vertex movement towards the center, and we have observed no cell neighbour swapping.
We checked that the features' center of mass had only a small displacement, which we subtracted from individual feature displacements without loss 
of generality.
To implement a quasi-1D analysis, we collected the features in a band
along the $Y$ axis and similarly along the $X$ axis  (Figure~\ref{fig:klt} (b-c)), 
and both axes were analysed separately.  The band
width (here $1/3$ of the initial circle diameter) was chosen to be large enough to perform large statistics on features, 
and small enough to treat the features displacement as one-dimensional.

\subsection{Comparison of model with experimental data}
We numerically solved 
adimensionalized
equations (see ESM)
using the Matlab solver \texttt{pdepde}. For a given set of parameter values $\left(\tau,\xi\right)$, we calculated $\varepsilon_0$ from the strain 
of the feature of largest initial amplitude, and rescaled
the solution of ESM Equation~21
so as to satisfy ESM Equation~10. 
Using the experimental initial positions of features
$\vec{R}_i^{\mathrm{exp}}(t = 0^+)$ and the calculated  strain field
$\varepsilon(z,t)$, we simulated the trajectories 
$\vec{R}_i^{\mathrm{sim}}(t)$,
$i = 1 \ldots N_f$, where $N_f$ was the number of features.

The mismatch $ {\cal E}$ with our computation was evaluated 
by comparison with the experimental trajectories, $\vec{R}_i^{\mathrm{exp}}(t) $:
\begin{equation}
  \label{eq:error}
  {\cal E} = \frac{1}{N_f} 
\sum_{i=1}^{N_f} \; \frac{1}{t_{\infty}} 
\sum_{t=1}^{t_{\infty}}  \; || \vec{R}_i^{\mathrm{sim}}(t) - \vec{R}_i^{\mathrm{exp}}(t) ||^2,
\end{equation}
where $|| \quad ||$ denoted the Euclidean norm, and
  $t_{\infty}$ the number of movie images.

\section*{Acknowledgments}
We gratefully thank F. Molino for modifying the ovuscule plugin, 
P. Th\'evenaz and S. Birchfield for advices regarding their softwares, H. Oda for reagents, J.-M. Allain for suggesting to use KLT algorithm, lab members for discussions, 
  and the PICT-IBiSA@BDD (UMR3215/U934) imaging facility of the Institut Curie.
This work was supported by grants to Y.B. from 
ARC (4830), ANR (BLAN07-3-207540), ERC Starting Grant (CePoDro 209718),
 CNRS, INSERM,  and Institut Curie; 
and by postdoc grants to I.B. by the FRM  (SPF20080512397), to F.B. by the NWO (825.08.033).

\newpage
\begin{widetext}

  \begin{center}
{\large    Electronic Supplementary Material (ESM)}
  \end{center}

 \setcounter{figure}{0}
 \setcounter{equation}{7}

\section*{Model}

The external dissipation is not accessible to single cell junction ablation 
experiments and its estimate requires modelling \cite{ESMgrill}. 
Our model is analogous to models of viscoelastic relaxation proposed
for local severing experiments of 
acto-myosin bundles in living cells \cite{Colombelli2009}.

 \subsection*{Hypotheses}

We use an Eulerian description. 
Since the severed domain area visibly decreases with time,
we do not enforce any in-plane incompressibility condition.
Note also that in-plane incompressibility would fix the velocity profile, which is incompatible with the experimental observation of different profiles (Figure~5).
We rather assume for simplicity that the out-of-plane thickness is a passive variable: hence the 3D Poisson ratio is $1/2$, but the in-plane effective 2D Poisson ratio vanishes. We test this hypothesis by checking {\it a posteriori}
that the model predictions are compatible with experimental observations.
Within this simplification, the
antero-posterior $X$ and medio-lateral $Y$   
axes (Figure~1) then  decouple.
Finally, we define $\sigma$ as the part of stress which has been removed by ablation.

We thus consider only one spatial dimension, with the variable $z$ denoting position along either
$X$ or $Y$.
We model the retracting slab of tissue as a linear viscoelastic Kelvin-Voigt material (that is, with a steady response at long time)
 with a Young modulus $E$ and a viscosity $\eta$.
We neglect the line tension around the tissue domain:
first, because we observe that the ellipse does not remain round (this suggests the line tension is weak, without being a proof, since we could also imagine that 
it has the same anisotropy as the strain); second, because even
 if the line tension exists, it is small enough that the ellipse contour remains visibly rough, 
see Movies 1-3.
Two terms contribute explicitly to $\sigma$: an elastic one
(which implicitly includes 
the external pressure) and a viscous one:
\begin{equation}
  \label{eq:stress}
\sigma(z,t) = 
E \; \varepsilon(z,t)  +   \eta \; \frac{\partial v}{\partial z}(z,t).
\end{equation}
Taking into account a
 fluid
external friction, with a friction coefficient {$\zeta$} of dimension 
Pa.s.m$^{-2}$,
the force balance equation reads:
\begin{equation}
  \label{eq:Newton}
 \frac{\partial \sigma}{\partial z}(z,t) - \zeta \;  v(z,t) = 0.
\end{equation}
Combining Equations~3, \ref{eq:stress} and \ref{eq:Newton} yields Equation~4.

\subsection*{Initial, boundary and final conditions}
The initial length of the retracting slab is $2 L$: 
$z \in [-L, L]$. Assuming that the initial state  respects mechanical equilibrium
(over the relevant time scale of 10 to 100~s),
 the initial 
strain is uniform, and we have for all $z$: 
\begin{eqnarray}
\varepsilon(z,t=0) &=& \varepsilon_0, \nonumber \\
 \sigma(z,t=0) &=& E \varepsilon_0.
  \label{eq:IC}
\end{eqnarray}
At the boundary of the severed region,  we have  for all $t$:
\begin{equation}
  \label{eq:BC:sigma}
  \sigma(\pm L,t) = 0.
\end{equation}
Since the final state is at mechanical equilibrium, we have for all $z$:
\begin{eqnarray}
 \lim\limits_{t \rightarrow \infty} \varepsilon(z,t) &= & 0, \nonumber \\
  \lim\limits_{t \rightarrow \infty} \sigma(z,t) &= &0.
\label{eq:fin:sigma}
\end{eqnarray}
Injecting Equation~3 into 
Equation~\ref{eq:stress}, using conditions given by Equations~\ref{eq:IC}-\ref{eq:fin:sigma} yields by integration over time:
\begin{equation}
  \label{eq:BC:e}
{  \varepsilon(\pm L,t) =  \varepsilon_0  \; e^{-t/\tau}},
\end{equation}
where the relaxation time is viscoelastic: $\tau = \frac{\eta}{E}$.

\subsection*{Initial velocity profile} 
Equations~3, \ref{eq:IC}, \ref{eq:BC:e} yield an approximate analytical prediction for the initial velocity profile at time $\delta t \ll \tau$, as follows.
An expansion of $\varepsilon$ at first order in $\frac{\delta t}{ \tau}$:
\begin{equation}
  \label{eq:e1:expand}
 \varepsilon(z,\delta t) = 
\varepsilon_0 +  \varepsilon_1(z) \; \frac{\delta t}{ \tau}, 
\end{equation}
defines the initial strain rate as:
\begin{equation}
  \label{eq:e1:def}
\frac{\varepsilon_1(z)}{\tau} = 
\frac{\varepsilon(z,\delta t) - \varepsilon_0}{\delta t}
\approx
\frac{\partial \varepsilon}{\partial t}(z,t=0) 
.
\end{equation}
Substituting Equations~\ref{eq:e1:expand}-\ref{eq:e1:def}
into Equation~4, we
obtain an ordinary differential equation for  $\varepsilon_1(z)$:
\begin{equation}
  \label{eq:e1:eqdiff}
  D \tau \; \frac{\mathrm{d}^2 \varepsilon_1}{\mathrm{d} z^2}  -
\varepsilon_1(z) = 0,
\end{equation}
with a diffusion coefficient  $D = \frac{E}{\zeta}$.
With the boundary condition
$\varepsilon_1(z = \pm L) = - \varepsilon_0$, the solution of 
Equation~\ref{eq:e1:eqdiff} reads:
\begin{equation}
  \label{eq:e1:sol}
  \varepsilon_1(z) =  - \varepsilon_0 
\; \frac{\cosh k z}{\cosh k L}
\end{equation}
where 
$k^2 = \frac{1}{D \tau}$, 
or $k L = \xi^{1/2}$ since $\xi = \frac{\zeta L^2}{\eta}$.
Spatial integration of Equation~\ref{eq:e1:sol} using Equations~3 and \ref{eq:e1:def} 
yields Equation~6.

\subsection*{Limit regimes}
ESM Figure~\ref{sfig:twolimits} shows two limit regimes of  Equation~4.

{\par \noindent $\bullet$ When internal viscosity dominates:} 
external friction may be neglected, and Equation~\ref{eq:Newton} reduces to
$
  \frac{\partial \sigma}{\partial z} = 0,
$
so that $\sigma$ is uniform and its value is fixed 
at the boundaries:
$\sigma(z,t) = \sigma(\pm L,t) = 0$ for all $(z,t)$.
Integration yields:
\begin{equation}
  \label{eq:relax:xi0}
  \varepsilon(z,t) =  \varepsilon_0 \; e^{-t/\tau}.
\end{equation}
The strain is uniform at all times
(ESM Figure~\ref{sfig:twolimits} (A)). 
Since the center of the slab does not move, $v(0,t) = 0$ for all $t$,
the velocity profile is linear:
\begin{equation}
  \label{eq:initv_modelfree}
v(z,t) =  \int_0^z  \frac{\partial \varepsilon}{\partial t}(z',t) \; \mathrm{d}z'
= - \frac{\varepsilon_0}{\tau}\;z
\; e^{-t/\tau}.
\end{equation}
As expected, in the limit of small friction $\xi \ll 1$,
Equation~6 reduces to Equation~\ref{eq:initv_modelfree}
when $t = 0$.
In this regime, all parts of the tissue relax exponentially with the same time scale $\tau$.

{\par \noindent $\bullet$  When external friction dominates:} the
internal viscosity of the tissue may be neglected. 
Combining Equations~3, \ref{eq:stress} and \ref{eq:Newton}
yields
a diffusion equation  for the strain field:
\begin{equation}
  \label{eq:diff:e}
 \frac{\partial \varepsilon}{\partial t} = D \;
\frac{\partial^2 \varepsilon}{\partial z^2}  
\end{equation}
with 
conditions given by Equations~\ref{eq:IC}-\ref{eq:fin:sigma}.
Strain diffuses from the boundaries on a time scale 
{ $\tau_D = L^2/D = \zeta L^2/E$}, and 
the strain field 
remains inhomogeneous until { $t \gg \tau_D$}
(ESM Figure~\ref{sfig:twolimits} (B)).

One may note that the strain diffusion equation (Equation~\ref{eq:diff:e}) 
is identical to that
obtained in the context of gel dynamics \cite{Charras2009,Doi2009}. Taking into
account interstitial flow ({\it i.e.} the transport of water across the
epithelium) would only modify the diffusion constant: 
$ D_{\textrm{permeation}}
 = E \left(
\frac{1}{\kappa} + \zeta \right)^{-1}$, where $\kappa$ is the permeability
coefficient.

\subsection*{Dimensionless variables}
 In the general case, to solve Equation~4 numerically we
   normalize variables by defining: { $\bar{z} = z/L$}, 
{ $\bar{t} = t/\tau$} and
{ $\bar{\varepsilon} = \varepsilon/\varepsilon_0$}.
Equation~4 then becomes: 
\begin{equation}
  \label{eq:general}
{ 
{\xi} \; \frac{\partial \bar{\varepsilon}}{\partial \bar{t}} = 
\frac{\partial^2 \bar{\varepsilon}}{\partial \bar{z}^2} +
\frac{\partial^3 \bar{\varepsilon}}{\partial \bar{t} \partial \bar{z}^2}}
\end{equation}
with the initial condition:
\begin{equation}
  \label{eq:IC:adim}
\bar{\varepsilon}(\bar{z},\bar{t}=0) = 1,
\end{equation}
the boundary conditions:
\begin{equation}
  \label{eq:BC:e:adim}
  \bar{\varepsilon}(\bar{z} = \pm 1,\bar{t}) =  e^{-\bar{t}},
\end{equation}
and the final conditions:
\begin{equation}
  \label{eq:final:e:adim}
  \bar{\varepsilon}(\bar{z},\bar{t}\rightarrow \infty) =  0,
\end{equation}

Since { ${\xi}  = \tau_D/\tau$}, 
the limit regimes are recovered when
{ $\tau_D \ll \tau$} (internal viscosity dominates, { ${\xi} \ll  1$}),
and   { $\tau_D \gg  \tau$} (external friction dominates,
{ ${\xi} \gg 1$}).

\newpage

\section*{ESM Figures}

\begin{figure}[ht]
\includegraphics[width=0.7\textwidth]{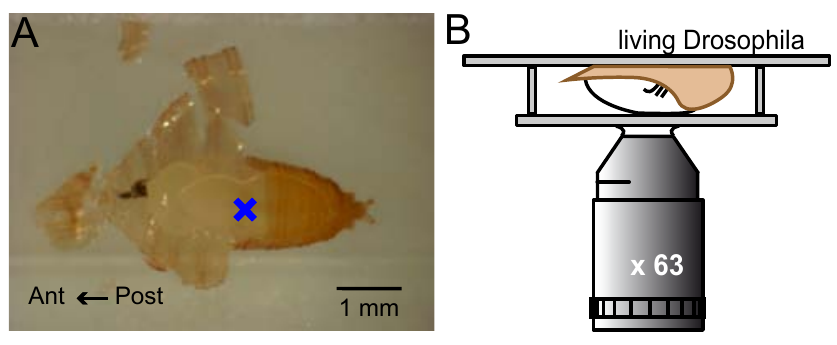}
\caption{\label{sfig:setup}
(ESM)
Set-up for  live imaging. (A) The  pupal case (light brown) is partially removed, while the cuticle (transparent) is kept intact. Blue cross: approximate position of severing, along the symmetry axis of the scutellum.
(B) The pupa (white) is mounted between slide and coverslip with its back facing the objective (dark grey).  
Antero-posterior axis is horizontal (anterior towards the left, posterior towards the right).}
\end{figure}

\begin{figure}[ht]
\includegraphics[width=0.7\textwidth]{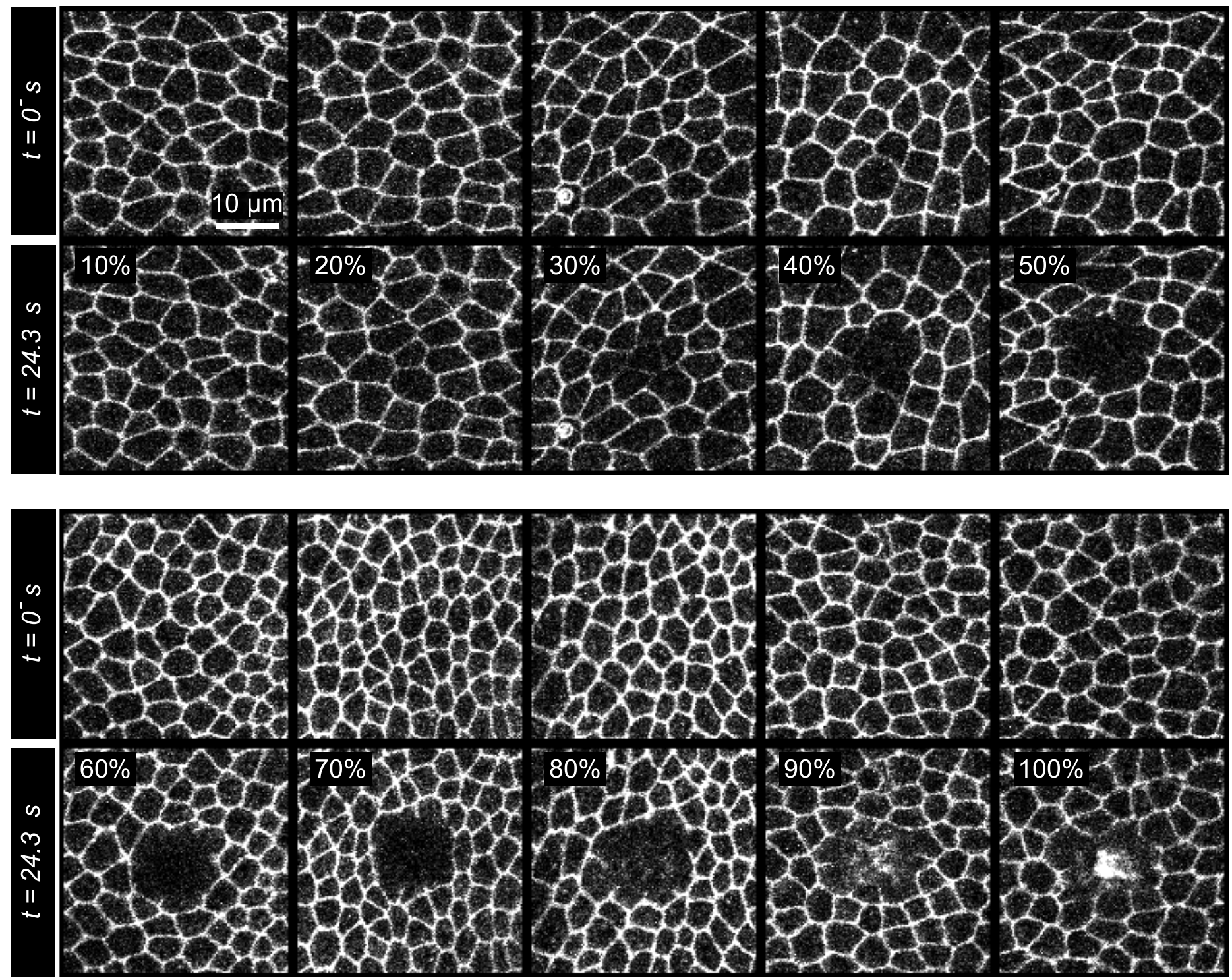}
\caption{\label{sfig:cavitation}
(ESM)
Range of acceptable severing powers. Images shown are just before the severing (top) and 24.3 s after the severing (bottom) as a function of the proportion (indicated at top left) of the full laser power, here at 890~nm. Below 20\%, no severing occurs.  From 30 to 50\%, severing is partial. From 60 to 80\%, severing is complete and cell-cell junctions move significantly; experiments 
 presented in this work  
are performed in this range.  Above 90\%, cavitation occurs. These ranges vary slightly from one experiment to the other, so that the quality of severing must be carefully assessed
{\it a posteriori}.}
\end{figure}

 \begin{figure}[ht]
\includegraphics[width=0.5\textwidth]{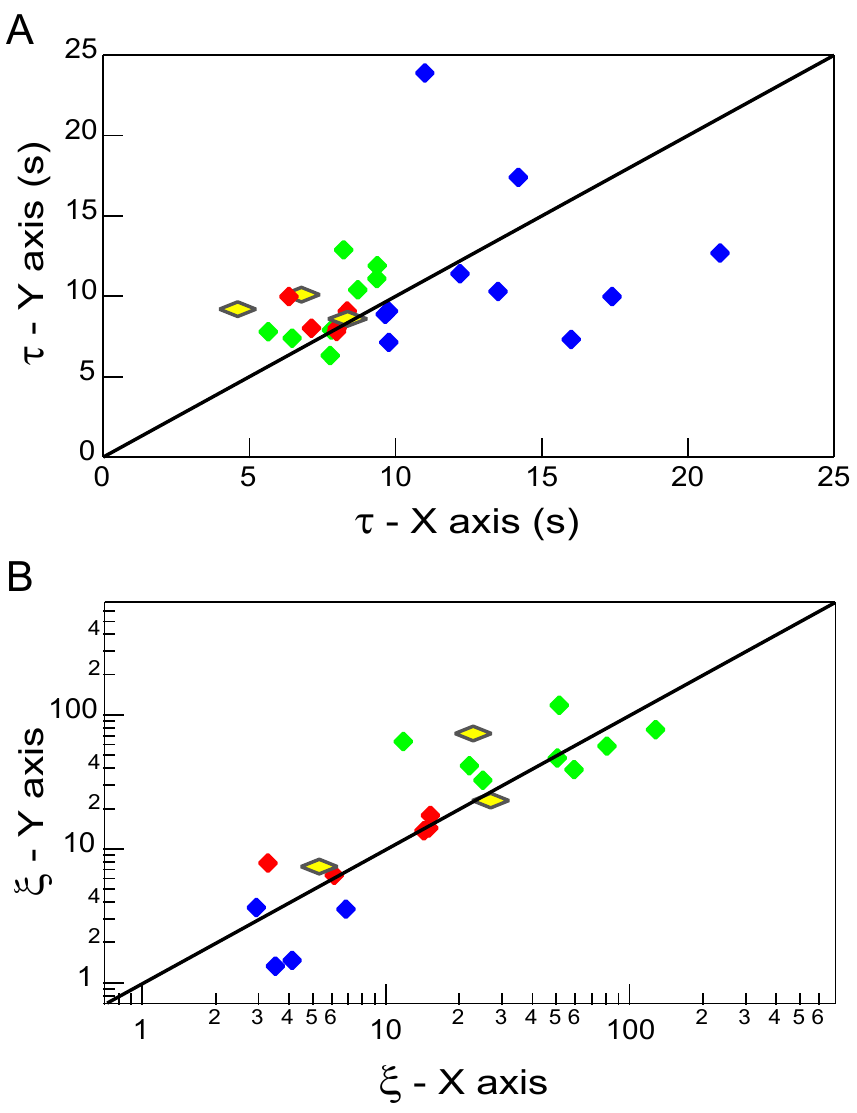}
\caption{ 
\label{sfig:xi_anisotropy}
 \label{sfig:tau_anisotropy}
(ESM)
  Check of $\tau$  and $\xi$ isotropy. 
(A) Relaxation time $\tau$:
  for each experiment, represented by one point, 
we plot the value obtained along the $Y$ direction 
(estimated as in the Inset of Figure~2)
{\it versus} the value obtained along the $X$ direction.
  Age is color-coded as in Figure~3: green, young; red,  middle-aged; blue, old. 
(B) Same  plot for   the friction/viscosity ratio $\xi$,
fitted by Equation~6 as in Figure~5. Note the logarithmic scale. 
 Two  values of $\xi$ of order  $10^{-3}$ and $10^{-4}$ are below the plotted range.
}
\end{figure}

\begin{figure}
\includegraphics[width=0.5\textwidth]{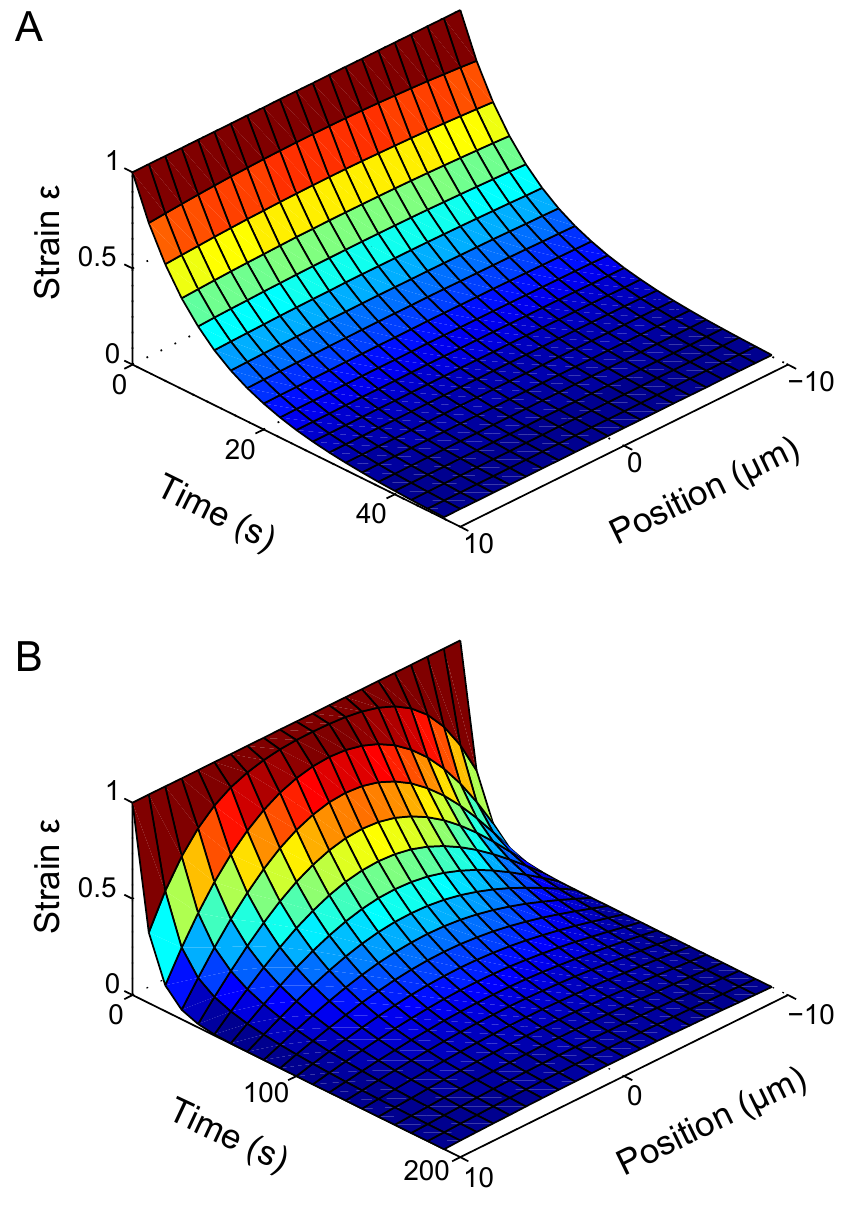}
\caption{(ESM)
Limit regimes: space-time profiles of strain fields obtained from numerical simulations
of Equation~4 for {$\varepsilon_0  = 1$}, {$\tau  = 10$ s}, {$L  = 10 \; \mu$m}.
(A) Uniform strain relaxation, when internal viscosity dominates (low $\xi$, initial velocity profile similar to the blue curve in Figure~5):  $D = 100 \; \mu \mathrm{m}^2.\mathrm{s}^{-1}$, 
$\tau_D = 1 \;s$.
(B) Strain diffusion, when external friction dominates (high $\xi$, initial velocity profile similar to the green curve in Figure~5):  $D = 1 \; \mu \mathrm{m}^2.\mathrm{s}^{-1}$, 
$\tau_D = 100 \;s$. Strain value in each grid square is color-coded from 0 (dark blue) to 1 (dark red). \label{sfig:twolimits}}
\end{figure}

 \begin{figure}
\includegraphics{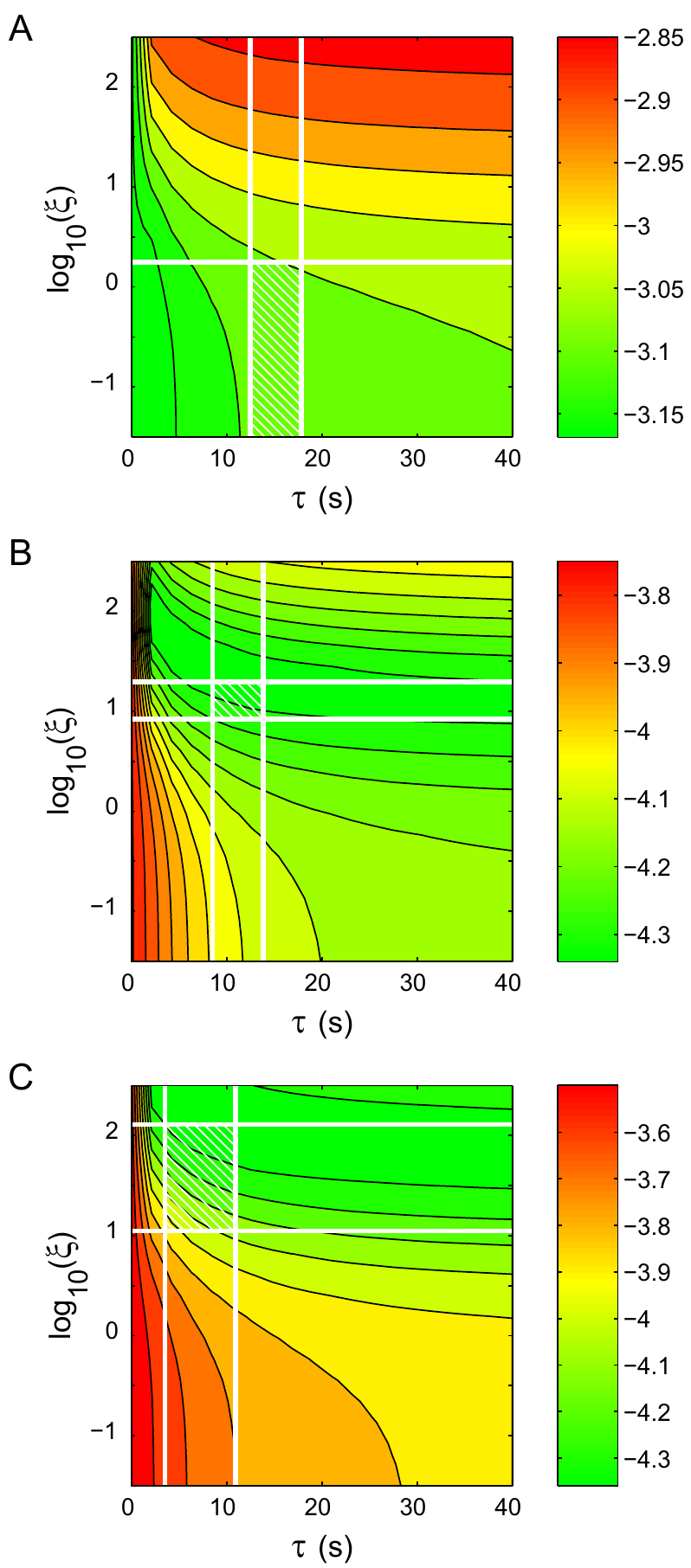}
\centering 
\caption{
\label{sfig:errormap} 
(ESM)
Error maps in the $(\tau,\xi)$ plane, semi-log axes. Experiments are those of 
Figure~5 and Movies 1-3; error is estimated according to Equation~7.
The minimal region of the error landscape (its deepest valley) is
too wide to allow a direct measurement of the parameters.
However, depending on the orientation of the minimal region 
of the error landscape, the computed error maps fall broadly 
into three categories, which agree with the three age groups.
There is also a good agreement with hatched regions, which are the results of the independent measurements of the parameters $\tau$ and $\xi$  
(Figure~3 (c) and ESM Figure~\ref{sfig:xi_anisotropy}).
(A)   Small $\xi$, old pupa (Movie 3).
Minimal valley parallel to the $\xi$ axis: this suggests that only an
upper bound, of the order of $1$ (white bar), can be given for the value of $\xi$.
Friction is negligible. 
(B)  Finite $\xi$,  middle-aged pupa (Movie 2). Minimal valley intermediate, parallel to the $\tau$ axis:
 friction cannot be neglected, 
and $\xi$ admits a finite estimate.
(C)  Large $\xi$, young pupa (Movie 1).  Minimal valley parallel to the $\tau$ axis, at very large $\xi$:
 this suggests that only a lower bound can be given 
for the values of $\tau$ and $\xi$.
}
\end{figure} 
 
\end{widetext}

\end{document}